\documentstyle[preprint,revtex]{aps}
\begin{document}
\draft
\preprint{IP/BBSR/92-85}
\begin{title}
Neutron matter - Quark matter phase\\
transition and Quark star
\end{title}
\author{H. Mishra, S.P. Misra, P.K. Panda}
\begin{instit}
Institute of Physics, Bhubaneswar-751005, India
\end{instit}
\author{B.K. Parida}
\begin{instit}
Physics Department, Regional College of Education,\\
Bhubaneswar-751007, India
\end{instit}
\begin{abstract}
We consider the neutron matter quark matter phase transition
along with possible existence of hybrid quark stars. The equation of state
for neutron matter is obtained using a nonperturbative method with pion
dressing of the neutron matter and an analysis similar to that of
symmetric nuclear matter. The quark matter sector is treated
perturbatively in the small distance domain.
For bag constant $B^{1/4}$=148 MeV, a first order phase transition is seen.
In the context of neutron quark hybrid stars, Tolman-Oppenheimer-Volkoff
equations are solved using  the equations  of state for quark matter and
for neutron matter with a phase transition as noted earlier.
Stable solutions for such stars are obtained with
the Chandrasekhar limit as
1.58 $M_\odot$ and radius around 10 km.
The bulk of the star is
quark matter with a thin crust of neutron matter of less than a
kilometer.
\end{abstract}
\pacs{}
\section {Introduction}
It is widely believed that neutron matter undergoes a phase
transition to quark matter at high densities and/or high
temperatures. The high temperature limit is expected
to have interesting consequences in heavy ion collision and/or
cosmology, whereas high baryon density behaviour is important for
the study of neutron stars.

The problem here for the treatment of nuclear matter or neutron matter
equation of state is basically nonperturbative.
A treatment of the same was developed by Walecka \cite{walecka}
which consists of
interactions of nucleons with a neutral scalar field $\sigma$ as well as a
neutral vector meson $\omega$.
These calculations however use meson fields as classical, and, use a
$\sigma $-field which is not observed.
An alternative model for infinite nuclear matter consisting of interacting
nucleons and pions was considered \cite{ijmpa,ijmpe} recently
where scalar isoscalar pion condensates
simulate \cite{deut,eisen} the effects of $\sigma$
mesons with  a phenomenological
short distance repulsion arising from composite structure of nucleons
or through vector meson
exchanges. We extend this nonperturbative technique to neutron matter
with the same parameters as for symmetric nuclear matter \cite{ijmpe},
and use this for hybrid stars.

We first consider neutron matter equation of state
at finite temperature
through thermofield dynamics \cite{tfd} as earlier \cite{ijmpe}. A
first order phase transition between neutron matter and quark matter
seems to be indicated. As stated, neutron matter is treated in a
nonperturbative manner. Quark matter is treated perturbatively
for high densities or short distances \cite{kapusta,ellis}.
Solutions of the Tolman-Oppenheimer-Volkoff (TOV) equations yield,  with  the
above equations of state, a hybrid star having a quark core and a crust
of neutron matter, the two states of matter being in equilibrium
at the interface \cite{glend}.  We obtain the Chandrasekhar limit for such
hybrid stars as 1.58 $M_{\odot}$

The outline of the paper is as follows.  In Sec. II we shall consider
neutron matter with pion condensates at zero temperature as well as
finite temperature.
In Sec. III we shall consider a standard form for quark matter with
pressure given in terms of chemical potential and discuss about the
phase transition from neutron matter to quark matter.
Section IV will consist of  calculations of the properties
of hybrid stars such as total mass, radius and moment of inertia.
In section V we summarise the results.

\section {Neutron matter equation of state}
We shall consider here neutron matter equation of state
at finite temperatures similar to the treatment of the same for symmetric
nuclear matter \cite{ijmpe}.
We start with the effective Hamiltonian for pion nucleon
interactions \cite{ijmpe,deut} as
\begin{equation}
{\cal H}_N(\vec x)={\cal H}_N^{(0)}(\vec x)+{\cal H}_{int}(\vec x)
\end{equation}
where the free nucleon part ${\cal H}_N^{(0)} $ is given as
\begin{equation}
{\cal H}^{(0)}_N(\vec x)=\psi_I^\dagger(\vec x) \epsilon_x \psi_I(\vec x)
\end{equation}
and the effective pion nucleon interaction part is given as
\begin{equation}
{\cal H}_{int}(\vec x)=\psi_I^\dagger(\vec x) \left[
-{iG\over 2 \epsilon_x }\vec \sigma.
\vec p \phi +{G^2\over 2 \epsilon_x }\phi^2\right]\psi_I(\vec x).
\end{equation}
In the above, $\epsilon _x=(M^2-\vec\bigtriangledown_x^2)^{1/2}$ where
$M$ denotes the nucleon mass.
Furthermore the free meson part of the Hamiltonian is
given as
\begin{equation}
{\cal H}_M(\vec x)={1\over 2}\left[{\dot \phi}_i^2
+(\vec \bigtriangledown \phi_i)\cdot(\vec \bigtriangledown \phi_i)
+m^2\phi_i^2\right]
\end{equation}
Clearly in the above $\psi_I$ refers to the large component of the
nucleon spinor and $\phi=\tau _i \phi_i$.

We expand the field operator $\phi_i(\vec z)$ in terms
of the creation and annihilation operators of off-mass shell mesons
satisfying equal time algebra as
\begin{equation}
\phi_i(\vec z)={1\over \sqrt{2 \omega _z}}(a_i(\vec z)^\dagger +a_i(\vec z))
\end{equation}
and
\begin{equation}
\dot\phi_i(\vec z)=i{\sqrt{\omega _z\over 2}}(a_i(\vec z)^\dagger -a_i(\vec z))
\end{equation}
where we take with the perturbative basis $\omega _z=(m^2-\vec \bigtriangledown
^2_z)^{1/2}$, with $m$ denoting
the mass of the meson.

We shall now consider finite temperature neutron matter. For this
purpose we shall use the methodology of thermofield dynamics \cite{tfd}.
We shall have the temperature dependent
background off-shell pion pair configuration given as
\begin{equation}
|f, \beta>=U_I(\beta)|f>=U_I(\beta)U|vac>
\end{equation}
where $U$ is given as
\begin{equation}
U =\exp \left({1\over 2}\int \tilde f(\vec k)a_i(\vec k)^\dagger a_i
(-\vec k)^\dagger d \vec k -h.c.\right)
\end{equation}
and $U_I(\beta )$ describes the effect of temperature.
The expression for this in terms of ordinary and thermal modes
is given as \cite{ijmpe,tfd}
\begin{equation}
U_I(\beta)=exp(B_I^\dagger(\beta )-B_I(\beta ))
\end{equation}
with
\begin{equation}
B_I(\beta )^\dagger={1 \over 2}\int \theta _B(\vec k, \beta )
b_i(\vec k)^\dagger\tilde b_i(-\vec k)^\dagger d\vec k
\end{equation}
In the above,
\begin{equation}
b_i(\vec k)^\dagger=Ua_i(\vec k)U^\dagger
\end{equation}
so that $b_i(\vec k)|f>=0$ and $\theta_B(\vec k,\beta)$ is a function
to be determined later. The parallel unitary transformation as in
Eq.(9) for the temperature dependance in neutron sector is given as
\cite{ijmpe}
\begin{equation}
U_{II}(\beta)=exp(B_{II}^\dagger(\beta )-B_{II}(\beta ))
\end{equation}
with
\begin{equation}
B_{II}(\beta )^\dagger={1\over 2}\int \theta _F(\vec k, \beta )
\psi_I(\vec k)^\dagger\tilde \psi_I(-\vec k)^\dagger d\vec k
\end{equation}
where $\theta _F(\vec k, \beta )$ will be determined later.
Thus we have the neutron matter density
\begin{eqnarray}
\rho (\beta )&=&<vac|U_{II}(\beta )^\dagger
\psi_ \alpha (\vec x)^\dagger \psi_ \alpha (\vec x)U_{II}
(\beta )|vac>\nonumber\\
& =&\gamma (2\pi)^{-3}\int d \vec k sin^2 \theta _F
\end{eqnarray}
where the degeneracy factor $\gamma=2$ for neutron matter \cite{ijmpe}.
$sin^2 \theta _F$ is the distribution function for the fermions. Clearly,
with $sin^2 \theta _F= \Theta(k_f-k)$, the step function, Eq. (14) gives
$\rho =\gamma k_f^3/6\pi^2$ at
zero temperature. For the interacting system $sin^2 \theta_F$
will be determined from the minimisation of the
thermodynamic potential, as is done latter.

We shall first calculate different contributions to the energy expectation
values corresponding to the Hamiltonian as in Eq. (1) and (4).
We thus have for the nucleon kinetic term
\begin{eqnarray}
h_f(\beta )&=& <vac|U_{II}(\beta )^\dagger \psi_I(\vec x)^\dagger
{(-\bigtriangledown_x^2)\over 2M}\psi_I(\vec x)U_{II}(\beta )|vac> \nonumber \\
&=&\gamma (2\pi)^{-3}\int d\vec k {k^2\over 2M}sin^2 \theta _F
\end{eqnarray}
The kinetic energy due to the mesons is given by
\begin{eqnarray}
h_k(\beta ) & = &<f,\beta|{\cal H}_M (\vec x)|f,\beta> \nonumber \\
& = & 3 (2\pi)^{-3}\int d\vec k \omega(\vec k)[sinh^2\tilde f(\vec k)
cosh2\theta_B(\vec k,\beta)+sinh^2\theta_B(\vec k,\beta)]
\label{eq:hk1}
\end{eqnarray}
where $\omega (\vec k)=\sqrt{\vec k^2+m^2}$.
We next derive the interaction energy density from Eq. (3) as \cite{ijmpe}
\begin{eqnarray}
h_{int}(\beta )&=& <f, \beta |U_{II}(\beta )^\dagger \psi_I(\vec x)^\dagger
\psi_I(\vec x){G^2\over 2 \epsilon _x}
\phi ^2(\vec x)U_{II}(\beta )|f, \beta > \nonumber \\
&\simeq & {G^2\over 2 M}\rho(\beta)<f,\beta|:\phi_i(\vec x)\phi_i(\vec x):
|f,\beta>\nonumber\\
&=& {G^2\rho(\beta )\over 2 M} I_{2M}
\label{eq:hint1}
\end{eqnarray}
where
\begin{equation}
I_{2M}={3\over(2\pi)^3}\int {d\vec k
\over \omega(\vec k)} \left({sinh2\tilde f(\vec k)cosh2\theta_B \over 2}
+sinh^2\tilde f(\vec k)cosh2 \theta _B+sinh^2 \theta _B\right).
\end{equation}
We shall now assume a phenomenological
term corresponding to meson repulsion due to composite structure of mesons
which is given as \cite{ijmpe}
\begin{equation}
h_m^R(\beta )=3a(2\pi)^{-3}\int \left(sinh^2\tilde f(\vec k)cosh2 \theta_B
+sinh^2 \theta _B\right) e^{R_\pi^2k^2} d\vec k,\label{eq:hmr2}
\end{equation}
where $a$ and $R_\pi$ are constants.
Finally, the nucleon repulsion term is taken as  \cite{ijmpe}
\begin{equation}
h_R=\lambda \rho^{2}(\beta)
\end{equation}
where $\rho(\beta)$ is as given in Eq. (14) and $\lambda$ is
another constant. This repulsion term could arise from
$\omega$-exchange or otherwise \cite{ijmpe}.

Finally, the energy density is given as
\begin{equation}
E(\beta)=(h_f(\beta)+h_m(\beta)+h_R(\beta))/\rho (\beta )
\end{equation}
where
\[h_m(\beta)=h_k(\beta)+h_m^R(\beta)+h_{int}(\beta).\]

The thermodynamic potential density $\Omega$ is given by \cite{fetter}
\begin{equation}
\Omega(\beta)=E(\beta)\rho -{\sigma\over \beta}-\mu_B\rho
\end{equation}
where the last term corresponds to nucleon number conservation with
$\mu_B $ as the chemical potential. We
may note that we shall be considering temperatures much below the nucleon
mass so that in the expression for $\rho(\beta)$ we do not include
antiparticle channel. The entropy density above is
$\sigma=\sigma_F+\sigma_B$ with $\sigma_F$ being the entropy in
fermion sector  given as \cite{tfd}
\begin{eqnarray*}
\sigma_F=-{\gamma \over (2\pi)^3}\int d\vec k &\Big[ &
sin^2 \theta _F(\vec k,\beta )
ln(sin^2 \theta _F(\vec k,\beta ))\\ &+&cos^2 \theta _F(\vec k,\beta )
ln(cos^2 \theta _F(\vec k,\beta ))\Big].\end{eqnarray*}
and similarly the meson sector contribution $\sigma_B$ is given as
\begin{eqnarray*}
\sigma_B=-{3\over (2\pi)^3}\int d\vec k &\Big[& sinh^2 \theta _B(\vec k,\beta )
ln(sinh^2 \theta _B(\vec k,\beta ))\\ &-& cosh^2 \theta _B(\vec k,\beta )
ln(cosh^2 \theta _B(\vec k,\beta ))\Big].\end{eqnarray*}
Thus the thermodynamic potential density $\Omega$ is now a functional of
$\theta_F(\vec k, \beta)$, $\theta_B(\vec k, \beta)$ as well as the
pion dressing function $\tilde f(\vec k)$ which will of course
depend upon temperature. Extremisation of Eq (21) with respect to
$\tilde f(\vec k)$ yields
\begin{equation}
tanh2\tilde f(\vec k)=-{G^2 \rho \over 2 M}\cdot {1\over {\omega ^2
(\vec k)+{G^2 \rho \over 2 M}+a \omega (\vec k)e^{R_\pi^2k^2}}}.
\end{equation}
Similarly minimising the thermodynamic
potential with respect to $\theta _B(\vec k, \beta )$ we get
\begin{equation}
sinh^2 \theta _B={1\over e^{\beta \omega ^\prime}-1}
\end{equation}
where
\begin{equation}
\omega ^\prime=(\omega +{G^2 \rho \over 2M \omega }+a e^{R_\pi^2k^2})
cosh2\tilde f(\vec k)+{G^2 \rho \over 2 M \omega}sinh2\tilde f(\vec k)
\end{equation}
Once we substitute the optimised dressing as in Eq. (23), the above
simplifies to
\begin{equation}
\omega ^\prime=(\omega +{G^2 \rho \over M \omega }+a e^{R_\pi^2k^2})^{1/2}
(\omega +a e^{R_\pi^2k^2})^{1/2}
\end{equation}
which is different from $\omega $ due to interactions.
Further, minimising the thermodynamic potential with respect to
$\theta _F(\vec k, \beta )$ we have the solution
\begin{equation}
sin^2 \theta _F={1\over e^{\beta (\epsilon_F-\mu_B)}+1}
\end{equation}
with
\begin{equation}
\epsilon_F={G^2\over 2M}I_{2M}+2 \lambda \rho +{k^2\over 2M}
\end{equation}
where $I_{2M}$ is given in equation (18). We may note that the change
in $\epsilon_F$ above from $k^2/2M$ is also due to interaction.

To calculate different thermodynamic quantities as functions of
baryon number density we first use Eq. (14) to calculate the chemical potential
$\mu_B$ in a self consistent manner with $\rho$ and
$\mu_B$ occuring also inside the
integrals through $sin^2\theta_F$ as in Eq. (27). Thus for each $\rho$,
we determine $\mu_B$ so that Eq. (14) is satisfied.
The ansatz functions $\theta _B$, $\theta _F$ and $\tilde f(\vec k)$
get determined analytically through the extremisation of the thermodynamic
potential. {\em The parameters $a$, $\lambda $ and $R_\pi$ are
fixed so as to reproduce the ground state properties of zero temperature
nuclear matter as is done earlier} \cite{ijmpe}. The parameter values
$\lambda=0.54$ fm$^2$, $R_\pi=1.18$ fm and $a=0.12$ GeV give the
nuclear matter single particle energy as $-15.03$ MeV at the
saturation density of $0.19$ fm$^{-3}$. We shall also use here the same
values for $a$, $R_\pi$ and $\lambda$.

With the thermodynamic potential determined
as above, we calculate different thermodynamic quantities.
Pressure is calculated using the thermodynamics relation \cite{fetter}
\begin{equation}
 P(\beta)=-\Omega (\beta)
\end{equation}
Pressure as a function of neutron matter density is plotted in Fig. 1
at temperatures of 0, 10, 15 and 20 MeV.
We note that as earlier \cite{ijmpa,ijmpe} the equation of state
is quite soft. In Fig 2 we plot pressure
versus baryon chemical potential at temperatures of 0, 5 and 10 MeV.
We note from Fig. 1 that at zero temperature pressure for neutron matter {\em
vanishes} at number densitiy of about 0.1/fm$^3$ so that the pressure at the
surface for a neutron star can be zero with a finite number density. This
unusual feature is due to the presence of nonperturbative contributions in
equation (28).

In the next section we shall consider quark matter as well as the transition
from neutron phase to quark phase. The
zero temperature version of equation (29) shall be used
for phase transition studies.
\section{Quark matter equation of state and phase transition}
Existence of quark matter in the core of neutron stars/pulsars is
an exciting possibility \cite{freedman,overgard,rosen}.
Densities of these stars
are expected to be high enough to force the hadron constituents or
nucleons to overlap thereby yielding quark matter. Since the
distance involved is small, perturbative QCD is used to derive
quark matter equation of state. We take the quark matter equation
of state as in Ref \cite{kapusta} in which u,d and s
quark degrees of freedom are included in addition to electrons.
As here \cite{kapusta} we set the electron, up and down
quark masses to zero and
the strange quark mass is taken to be 180 MeV. In chemical
equilibrium $\mu_d=\mu_s=\mu_u+\mu_e$. In terms of baryon and
electric charge chemical potentials $\mu_B$ and $\mu_E$, one has
\begin{equation}
\mu_u={1\over 3}\mu_B+{2\over 3}\mu_E,\quad
\mu_d={1\over 3}\mu_B-{1\over 3}\mu_E,\quad
\mu_s={1\over 3}\mu_B-{1\over 3}\mu_E.
\end{equation}

The  pressure contributed by the quarks is computed to order
$\alpha_s=g^2/4\pi$ where $g$ is the QCD coupling constant.
Confinement is simulated by a bag constant B. The electron pressure is
\cite{kapusta}
\begin{equation}
P_e={\mu_e^4\over 12\pi^2}.
\end{equation}
The pressure for quark flavor f, with f=u,d or s is \cite{kapusta}
\begin{eqnarray}
P_f&=&{1\over 4\pi^2}\left[\mu_f k_f(\mu_f^2-2.5m_f^2)+
1.5m_f^4ln\left({\mu_f+k_f\over m_f}\right)\right]\nonumber\\
&-&{\alpha_s\over \pi^3}\left[ {3\over 2}\left(\mu_f k_f-m_f^2
ln\left({\mu_f+k_f\over m_f}\right)\right)^2-k_f^4\right].
\end{eqnarray}
The Fermi momentum is $k_f=(\mu_f^2-m_f^2)^{1/2}$. The total pressure,
including the bag constant B, is
\begin{equation}
P=P_e+\sum_f P_f-B.
\end{equation}
There are only two independent chemical potentials $\mu_B$ and
$\mu_E$. $\mu_E$ is adjusted so that the matter is electrically
neutral, i.e. $\partial P/\partial \mu_E=0$. The baryon number density
is given by $\rho=\partial P/\partial \mu_B$.

We now consider the scenario of phase transition from cold neutron
matter to quark matter. As usual, the phase boundary of the
coexistence region between the neutron and quark phase is determined
by the Gibbs criteria. The critical pressure and critical chemical potential
are determined by the condition
\begin{equation}
P_{nm}(\mu_B)=P_{qm}(\mu_B).
\end{equation}
The r.h.s of equation (34) is the same as P in equation (33), where as
the l.h.s is the zero temperature limit of equation (29).
We take $\alpha_s=0.4$ and the bag constant $B=(148\;MeV)^4$,
which is a reasonable
value to calculate pressure in the quark sector.
In Fig 3 we plot pressure versus chemical potential $\mu_B$
for cold neutron matter
and quark matter as considered here. The $(P,\mu)$ curves for quark matter and
neutron matter yield the critical parameters as
$P_{cr}=4$ MeV/fm$^3$ and $(\mu_B)_{cr}=960$ MeV.
The corresponding critical energy densities for neutron matter and
quark matter respectively are given as $\epsilon_{cr}^{nm}=226$ MeV/fm$^3$ and
$\epsilon_{cr}^{qm}=284$ MeV/fm$^3$ . The corresponding baryon number
densities are $\rho_B^{nm}=0.24$/fm$^3$ and $\rho_B^{qm}=0.33$/fm$^3$.
Thus there is a first order phase transition. We also note that the phase
transition seems to occur around number density of only
about one and half times the
nuclear matter density. There is a second crossing point in $(P,\mu)$ curves
not seen in Fig.3.
However, it occurs at higher densities where we do not
expect neutrons to exist because of overlap of quark wave functions
so that neutron matter does not exist.

The phase transition as above was possible due to the first two
terms in equation (28). These terms arise in a nontrivial manner
from interactions, with $I_{2M}$ given in equation in (18). We note
that here we are using the same parameters for neutron matter as
earlier for symmetric nuclear matter, giving rise to a soft
equation of state \cite{ijmpe}.

The early phase transition from neutron matter to quark matter
obviously implies that the interior of ``neutron star" will
usually consists of quark matter. We investigate this possibility
in the next section.
\section{Hybrid stars}
For the description of neutron star, which is highly concentrated
matter so that the metric of space-time geometry is curved, one has to
apply Einstein's general theory of relativity. The space-time
geometry of a spherical neutron star described by a metric in
Schwarzschild coordinates has the form \cite{weinberg,weber}
\begin{equation}
ds^2=-e^{\nu(r)}dt^2+[1-2M(r)/r]^{-1}dr^2+r^2[d\Theta^2+sin^2 \Theta
d\phi^2]
\end{equation}
The equations which determine the star structure and the geometry are,
in dimensionless forms \cite{weinberg}
\begin{mathletters}
\begin{equation}
{d\hat P(\hat r r_0)\over d\hat r}=-\hat G
{[\hat\epsilon (\hat r r_0)+\hat P (\hat r r_0)][\hat M (\hat r r_0)
+4\pi a \hat r^3 \hat P(\hat r r_0)]\over \hat r^2[1-2\hat G
\hat M (\hat r r_0)/\hat r]},\label{tov1}
\end{equation}
\begin{equation}
\hat M (\hat r r_0)=4\pi a \int_0^{\hat r r_0} d\hat r^\prime
\hat r^{\prime^2} \hat \epsilon(\hat r^\prime r_0),\label{tov2}
\end{equation}
and the metric function, $\nu (r)$ is given by
\begin{equation}
{d\nu(\hat r r_0)\over d\hat r}=2\hat G {[\hat M (\hat r r_0)
+4\pi a \hat r^3 \hat P(\hat r r_0)]\over \hat r^2[1-2\hat G
\hat M (\hat r r_0)/\hat r]}.\label{tov3}
\end{equation}
\end{mathletters}
In equations (36) the following substitutions have been made.
\begin{mathletters}
\begin{equation}
\hat \epsilon\equiv \epsilon/\epsilon_c,\quad
\hat P\equiv P/\epsilon_c,\quad\hat r\equiv r/r_0,\quad
\hat M\equiv M/M_\odot,
\label{tov4}
\end{equation}
where, with
$ f_1=197.329 $ MeV fm and $r_0=3\times 10^{19}$ fm,
we have
\begin{equation}
a\equiv\epsilon_c r_0^3/M_\odot, \quad
\hat G\equiv {G f_1 M_\odot\over r_0}
\label{tov5}
\end{equation}
\end{mathletters}
In the above, quantities with hats are dimentionless.
G in equation (\ref{tov5}) denotes the gravitational constant with
$G=6.707934\times 10^{-45}\;\mbox{MeV}^{-2}$.

In order to construct a stellar model, one has to integrate equations
(\ref{tov1}) to  (\ref{tov3}) from the star's center at r=0 with
a given central energy density $\epsilon_c$ as input until the
pressure $P(r)$ at the surface vanishes. As stated  in the last section,
with any reasonable
central density, we expect that at the center
 we shall have quark matter, and not neutron matter.
Hence we shall be using here
the equation of state for quark matter through equations (32) and (33) with
$\hat P(0)=P(\epsilon_c)$. We then integrate the TOV equations
until the pressure
and density decrease to their critical values, so that there is a first
order phase transition from quark matter to neutron matter at radius
$r=r_c$. For $r>r_c$, we shall have equation
of state for neutron matter where  pressure will change continuously
but the energy density will have a discontinuity at $r=r_c$.
TOV equations with equation
of state for neutron matter shall be continued until we reach the density
around 0.1 particles/fm$^3$ when the pressure vanishes (see Fig 1).
This will complete the calculations for stellar
model for hybrid ``neutron" star,
whose mass and radius can be calculated for different central densities.

We now plot in Fig 4 the mass of a star as a function of central
energy density to examine the stability of such stars. As may be seen from
the figure $dM/d\epsilon_c$ becomes negative around 1540 MeV/fm$^3$
after which they become unstable and may collapse into black holes
with  the Chandrasekhar limit as 1.58$M_{\odot}$
\cite{weinberg,shapiro}. The dashed part of the curve indicates the
instability region.
Fig 5 shows the mass as a function of radius  for such stars obtained
for different central densities varying in the range
900 to 1500 MeV/fm$^3$. This yields stable hybrid stars of masses $M\simeq$
1.49 to 1.58 $M_\odot$ with raddi $R\simeq$ 9.85 to 9.4 km respectively,
similar to the results of Ref. \cite{rosen}.

The energy density profile obtained from (\ref{tov1}) to (\ref{tov3}) are
plotted in Fig 6 for central densities $\epsilon_c=900$ MeV/fm$^3$
and $\epsilon_c=1500$ MeV/fm$^3$.
As we go away from the core towards the surface through TOV
equations, when the critical pressure is reached, the density
drops discontinuously indicating the first order phase transition.
 Thus e.g. for central density of 900 MeV/fm$^3$ such a star has a quark
matter core of radius 9 km with a crust of neutron matter of about 0.85 km,
whereas for $\epsilon_c$=1500 MeV/fm$^3$, the quark matter core radius is
8.9 km with neutron matter crust of 0.5 km.

It may be interesting to consider smaller central densities, when neutron
matter is expected to be more abundant. For example, we have noted that
at central densities of $\epsilon_c$=700 Mev/fm$^3$ to
350 Mev/fm$^3$, $R$ changes from 9.87 km to 7.3 km, whereas the quark
core changes from 9 km to 5.1 km. The neutron crust here increases from
0.87 km to 2.2 km, and the mass of the star decreases to as low as
$0.32M_\odot$. It is not clear to us whether, such small hybrid
stars or for that matter neutron stars can get formed.

We may also calculate the surface gravitational red shift $Z_s$ of
photons to see the possibility of distinguishing such a hybrid star
from a neutron star or a quark star. It is given by \cite{brecher}
\begin{equation}
Z_s={1 \over\sqrt{[1-2GM/R]}} -1.\label{zs}
\end{equation}
In Fig 7 we plot $Z_s$ as a function of $M/M_{\odot}$.
As may be seen, $Z_s$ as related to mass and radii of hybrid stars is
similar to that of pure neutron stars and hence a measurement of the same
will not be an evidence for the existance of such a star.
It is however possible that the discrete slowing down of pulsars
due to the presence of two states of matter with various radii
shall throw some light on the above structure. For a quantitative
estimate of the same, however, we shall need interactions at the
interface, which at present looks impossible to tackle. This will
need for example the expression for the moment of
inertia of pulsars as well as the equillibrium for the rotating
hybrid stars given as \cite{hartle}
\begin{equation}
I={8\pi\over 3}\epsilon_c r_0^5\int_0^{\hat R r_0}d {\hat r}\hat r^4
{[\hat\epsilon (\hat r r_0)+\hat P (\hat r r_0)]e^{-\nu(\hat r r_0)/2}
\over\sqrt{[1-2\hat G \hat M (\hat r r_0)/\hat r]}}.\label{momi}
\end{equation}

We may further take the relativistic Keplerian angular velocity
$\Omega_k$ to be given as \cite{friedman}
\begin{equation}
\frac{\Omega_k}{10^4 s^{-1}}=0.72\sqrt{{M/M_\odot\over (R/10km)^3}}
\end{equation}
as for neutron stars and estimate the same.
Fig 8 shows the variation of $ {\Omega_k}$
as a function of $M/M_{\odot}$ for
such a star.

In Fig 9 we plot the fraction of the quark matter in a star as a
function of the central density. As can be seen from the figure
quark matter constitutes the dominant contribution to the mass of such stars.
The internal
quark matter structure of such hybrid stars could influence neutrino
emissivity due to larger number of $\beta$-decay channels in quark matter
phase which may be relevant for the dynamics of supernova explosions
giving rise to pulsars, now expected to be really hybrid stars.

\section{conclusions}
Let us summarise the findings of the present paper.

Using a $nonperturbative$ equation of state for neutron matter and a usual
{\em perturbative} equation of state for quark matter we saw that a
first order phase transition exists between the neutron phase and quark phase.
We note that neutron matter equation of state practically has no free
parameters
since the parameters are fixed \cite{ijmpe} to yield the standard nuclear
matter
properties. Quark matter equation of state has the parameters : $m_s$,
$\alpha_s$ and B. The bag constant B, turns out to be a very sensitive
parameter. We found, for example, for $\alpha_s=0.4$, neutron matter quark
matter
transition occurs for $B\simeq (148 MeV)^4$. As can be seen in Fig. 3 there
is not much freedom for B.

The phase transition from neutron matter to quark matter seems to
indicate that the core of a ``neutron star" shall consist of quark matter.
To study the hybrid star, namely a star consisting of both
quark matter and neutron matter, we applied the TOV equation to the appropriate
equations of state with a given central energy density $\epsilon_c$.
It turns out that stable hybrid stars with a core of quark matter
and a  crust of neutron
matter can exist upto $\epsilon_c\simeq 1500$ MeV/fm$^3$ beyond
which instability is indicated.
The bulk of the hybrid star is provided by the quark matter,
the neutron matter providing only a thin crust. It is thus the quark
matter equation of state and the corresponding parameters which shall play
a dominant role in the formation of
hybrid stars. Pulsars are expected to be stars of this type, but the
gross properties appear to be similar to what we believe regarding neutron
stars. We may recall that the nuclear matter equation of
state used here is rather soft with compressibility
of about 150 MeV.

We have also calculated other gross properties of such hybrid stars like
moment of inertia, surface gravitational red shift and Keplerian rotation
period. The results are similar to those obtained by others \cite{rosen}.
\hfil\break
\acknowledgments
The authors are thankful to Somenath Chakrabarty,
A. Mishra, S.N. Nayak and Snigdha Mishra for
many discussions. SPM wishes to thank the Department of Science and
Technology, Government of India for the research grant
SP/S2/K-45/89 for financial assistance.

\figure{Pressure P as a function of neutron matter density $\rho$
is plotted for
temperatures T=0 MeV, 10 MeV, 15 MeV and 20 MeV.}
\figure{Pressure P of neutron matter as a function of
baryon chemical potential is plotted for
temperatures T=0 MeV, 5 MeV and 10 MeV.}
\figure{We plot here pressure versus baryon chemical potential
for neutron matter and for quark matter with $\alpha_s$=0.4
and $B^{1/4}$ =148 MeV at zero temperature.
The critical parameters may be seen as
$P_{cr}$=4 MeV/fm$^3$ and $(\mu_B)_{cr}$=960 MeV.}
\figure{ We plot mass of the hybrid star as a function of
central density. The instability region is shown as dashed
curve.}
\figure{We plot mass as a function of radius for hybrid stars here.
The dashed line here indicates the instability of the solutions.}
\figure{We plot here energy density profile for central densities
$\epsilon_c$=900 MeV/fm$^3$ and $\epsilon_c$=1500 MeV/fm$^3$.
The discontinuities
at the ends indicate a first order phase transition
at the interface of quark matter and neutron matter.}
\figure{We plot here surface gravitational red shift $Z_s$ versus
the star mass.}
\figure{We plot here Keplerian angular velocity $\Omega_k$
versus star mass.}
\figure{We plot here the ratio between the mass in the quark matter
phase and the total mass of the star as a function of its central density.
The dashed curve indicates the instability region.}

\begin{references}
\bibitem{walecka}  J.D. Walecka, Ann. Phys. {\bf 83} (1974) 491; B.D. Serot
and J.D. Walecka, Adv. Nucl. Phys. {\bf 16} (1986) 1; S. Gmuca,
J. Phys.  {\bf G17}, (1991) 1115.
\bibitem{ijmpa}  A. Mishra, H. Mishra and S.P. Misra, Int. J. Mod. Phys.
{\bf A7} (1990) 3391.
\bibitem{ijmpe} H. Mishra, S.P. Misra, P.K. Panda and B.K. Parida,
to appear in Int. J. Mod. Phys. {\bf E}.
\bibitem{deut}P.K. Panda, S.P. Misra and R. Sahu, Phys. ReV. {\bf C45}
(1992), 2079.
\bibitem{eisen}  J.M. Eisenberg, Phys. Lett B {\bf 104} (1988) 353;
H. Jung, F. Beck and G.A. Miller, Phys. Rev. Lett. {\bf 62} (1989) 2357.
\bibitem{tfd}  H. Umezawa, H. Matsumoto and M. Tachiki {\it Thermofield
dynamics and condensed states} (North Holand, Amsterdam, 1982)
\bibitem{kapusta} J.I. Kapusta, {\it Finite Temperature Field Theory}
(Cambridge University Press)
\bibitem{ellis} J. Ellis, J.I. Kapusta and K.A. Olive, Nucl. Phys. {\bf B348}
(1991) 345.
\bibitem{glend} N. K. Glendenning, Talk presented at the Workshop
on Strange Quark Matter in Physics and Astrophysics,
Aarhus, Denmark, 1991, to appear in the Proceedings; ibid,
Preprint LBL-30878.
\bibitem{fetter} See for example in {\it Quantum Theory of Many Particle
System} A.L. Fetter and J.D. Walecka, McGraw Hill Book
Company, 1971.
\bibitem{freedman} B. Freedman and L. McLerran, Phys. Rev. {\bf D17}
(1978) 1109; B.D. Serot and H. Uechi, Ann. Phys. {\bf 179} (1987) 272.
\bibitem{overgard} T. \O verg\.{a}rd and E. \O verg\.{a}rd, Class.
Quantum Grav. {\bf 8} (1991) L49.
\bibitem{rosen} A. Rosenhauer et al  Nucl. Phys. {\bf A540} (1992) 630.
\bibitem{weinberg} S. Weinberg, {\it Gravitation and cosmology}, (Wiley,
New York, 1972).
\bibitem{weber} F. Weber and M.K. Weigel, Nucl. Phys. {\bf A493} (1989) 549.
\bibitem{shapiro}S. L. Shapiro and S. A. Teulosky, {\it Black holes,
white dwarfs and neutron stars} (Wiely, New York, 1983).
\bibitem{brecher}K. Brecher, Astro. Phys. J. {\bf 215} (1977) L17.
\bibitem{hartle} J.B. Hartle, Phys. Reports {\bf 46} (1978) 201.
\bibitem{friedman}J. L. Friedman, J.R. Ipser and L. Parker, Phys. Rev. Lett.
{\bf 62} (1989) 3015.
\end{references}
\end{document}